\documentclass[a4paper,aps,prd,twocolumn,groupedaddress,nofootinbib,floatfix]{revtex4}
\usepackage{aas_macros}

\usepackage{mathptmx}
\usepackage{helvet}
\usepackage{courier}

\usepackage[T1]{fontenc}

\begin{document}

\title{Reply to ``Comment on `Inference with minimal Gibbs free energy
in information field theory' '' by Iatsenko, Stefanovska and McClintock}
\author{Torsten A. En{\ss}lin}
\affiliation{Max-Planck-Institut f\"ur Astrophysik, Karl-Schwarzschild-Str. 1, 85741 Garching, Germany}
\author{Cornelius Weig}
\affiliation{Arnold Sommerfeld Center for Theoretical Physics and Center for NanoScience, Department of Physics, Ludwig-Maximilians-Universit{\"a}t M{\"u}nchen, Theresienstra{\ss}e 37, 80333 M{\"u}nchen, Germany}
\date{\today}
\begin{abstract}
We endorse the comment on our recent paper [{En{\ss}lin} and {Weig}, {\pre} 82,  051112 (2010)]
by Iatsenko, Stefanovska and McClintock [Phys. Rev. E 85 033101 (2012)]
and we try to clarify the origin of the apparent controversy on two issues.
The aim of the minimal Gibbs free energy approach to provide a signal
estimate is not affected by their Comment. However, if one wants to
extend the method to also infer the a posteriori signal uncertainty
any tempering of the posterior has to be undone at the end of the
calculations, as they correctly point out.
Furthermore, a distinction is made here between \textit{maximum entropy}, the \textit{maximum
entropy principle}, and the so-called \textit{maximum entropy method}
in imaging, hopefully clarifying further the second issue
of their Comment paper.
\end{abstract}
\maketitle
Iatsenko, Stefanovska and McClintock \citep{Iatsenkocomment2012}
point out two possible sources of confusion about how the minimal
Gibbs free energy method which was introduced in our
paper \citep{2010PhRvE..82e1112E}, has to be used.

The first point of confusion concerns the usage of temperature in the Gaussian approximation
of a high dimensional probability density function. This was introduced
to enable us to emphasize regions near (temperature $T<1)$ or more distant
from ($T>1$) the maximum by giving them a larger weight in the approximation.
Our original claim is still correct in that the signal mean can be
read off approximately from the mean of this tempered Gaussian. If the uncertainty information is
 also extracted from the
tempered Gaussian, the tempering has to be undone at the end of the
calculation, as was pointed out correctly in \citep{Iatsenkocomment2012}.
This was actually implicitly done in our paper in Eq. (44) and
very indirectly indicated in the sentence enclosing this equation.
However, we fully agree with \citep{Iatsenkocomment2012} that this
important fact should have been spelled out more prominently.

The second point is the slightly deceptive usage of the term \textit{maximum
entropy} (ME) in our paper, which is defined there as the unconstrained
maximum of the Boltzmann/Shannon entropy. This term should not be
confused with the \textit{maximum entropy principle} (MEP) \citep{1957PhRv..106..620J},
which is the constrained maximization of such an entropy, a distinction
correctly made by \citep{Iatsenkocomment2012}. Our aim in using the
terminology of ME was to highlight the importance of incorporating
phase-space volume factors correctly into inference schemes, which
counteracts the attraction of the posterior maximum and thereby impedes
over-fitting. 

We might have given the impression that we were arguing against the usefulness
of the MEP by discussing critically, but only briefly, the often used
\textit{maximum entropy method} (MEM) in imaging. In hindsight,
it might have been wiser to put this discussion into a separate publication, so as
not to risk further confusion. In any case, we want to explain
the difference here.

The MEM imaging can be derived from the MEP in the specific case in which the image
is composed of a nearly infinite number of uncorrelated and in principle
distinguishable flux elements of an infinitesimally small strength each
\citep{1972JOSA...62..511F}. In this case, each possible image can
be attributed to an internal image entropy, due to the number of ways
the flux elements can be redistributed among pixels without changing
the image. MEM can be shown to be a maximum a posteriori method, with
the internal entropy prior suppressing large flux values stronger
than exponentially \citep{2010PhRvE..82e1112E}. Although the conditions
under which MEM is optimal are rather artificial and rarely met for
the problem at hand, its application is widespread. However, ME and
the MEP are more general, in that they not only deal with the internal
entropy of a single image realization, if this can be defined, but
with the full entropy of the phase space of all possible images. The
discussion of this issue in our paper might have given the impression
that we wanted to argue against the MEP, while we merely criticized the unjustified
usage of MEM image restoration. We want to emphasize here, that we
consider the MEP to be a correct and important concept.
\vspace{-3em}
\begin{acknowledgements}
We thank D. Iatsenko, A. Stefanovska and P. V. E. McClintock for the collegial way they discussed their comments with us.
\end{acknowledgements}
\bibliography{../Bib/ift}
\end{document}